\documentclass[aps,prd,preprint,nofootinbib,superscriptaddress]{revtex4}
\usepackage{amsmath,amssymb,amsbsy,color}
\usepackage{graphicx}
\usepackage{psfrag}
\usepackage{graphicx}

\newcommand{\nn}{\nonumber}

\newcommand{\pslash}{p\kern-1ex /}
\newcommand{\lslash}{l\kern-1ex /}
\newcommand{\sslash}{s\kern-1ex /}
\newcommand{\Dslash}{{\cal D}\kern-1.5ex /}

\newcommand{\tr}{{\rm tr}}

\newcommand{\beqa}{\begin{eqnarray}}
\newcommand{\eeqa}{\end{eqnarray}}

\begin{document}
\preprint{UTHEP-547}
\preprint{RIKEN-TH-105}
\preprint{KEK-CP-195}
\preprint{YITP-07-41}
\title{Finite volume QCD at fixed topological charge}

\author{Sinya Aoki}
\affiliation{
  Graduate School of Pure and Applied Sciences, 
  University of Tsukuba, Tsukuba 305-8571, Ibaraki Japan}
\affiliation{
  Riken BNL Research Center, 
  Brookhaven National Laboratory, Upton, NY 11973, USA}

\author{Hidenori Fukaya}
\affiliation{
  Theoretical Physics Laboratory, RIKEN, Wako 351-0198, Japan}

\author{Shoji Hashimoto}
\affiliation{
  High Energy Accelerator Research Organization (KEK), 
  Tsukuba 305-0801, Japan}
\affiliation{
  School of High Energy Accelerator Science,
  The Graduate University for Advanced Studies (Sokendai),
  Tsukuba 305-0801, Japan}

\author{Tetsuya Onogi}
\affiliation{
  Yukawa Institute for Theoretical Physics, 
  Kyoto University, Kyoto 606-8502, Japan}

\date{\today}

\begin{abstract}
  In finite volume the partition function of QCD with a given $\theta$ is
  a sum of different topological sectors with a weight primarily
  determined by the topological susceptibility.
  If a physical observable is evaluated only in a fixed topological sector,
  the result deviates from the true expectation value by an amount
  proportional to the inverse space-time volume $1/V$.
  Using the saddle point expansion, we derive formulas to express the
  correction due to the fixed topological charge in terms of a $1/V$
  expansion. 
  Applying this formula, we propose a class of methods to determine
  the topological susceptibility in QCD from various correlation functions
  calculated in a fixed topological sector.
\end{abstract}

\maketitle

\section{Introduction}

Quantum Chromodynamics (QCD) in four space-time dimensions allows topologically
non-trivial gauge configurations labeled by a winding number or a topological
charge $Q$.
The path integral to define the partition function of QCD includes an
integral over configurations with arbitrary $Q$.
In order to ensure the cluster decomposition property of physical observables,
the weight among different topological sectors must be 
$e^{i\theta Q}$, which defines the $\theta$ vacuum of QCD.
If one considers a path integral restricted in a fixed topological sector, the
cluster decomposition property ---one of the necessary properties of quantum
field theory--- is violated \cite{Weinberg}. 
In this paper we address the question of fixing topology in the context of
non-perturbative calculation of QCD on the lattice.
But the analysis does not depend on any particular regularization of the
theory. 

In the lattice QCD simulations the inclusion of the effects of dynamical quarks
is computationally most demanding.
Since the direct computation of the fermion determinant $\det(D+m)^2$, with
$D$ the Dirac operator on the lattice and $m$ the quark mass, requires
prohibitive computational cost, one usually introduces a pseudo-fermion field
$\phi$ to write the determinant in the form
$\int[d\phi][d\phi^\dagger]\exp(-\phi^\dagger(D+m)^{-2}\phi)$,
so that the problem is reduced to an evaluation of the inverse fermion matrix
$(D+m)^{-1}$.
Since the effective lattice action becomes non-local, the Monte Carlo updation
is most efficiently done by updating all the gauge links on the lattice at
the same time introducing a molecular dynamics evolution.
The most popular such algorithm to date is the Hybrid Monte Carlo algorithm
\cite{Duane:1987de} that combines the molecular dynamics evolution with a
Metropolis accept/reject step. 

With the molecular dynamics evolution, the ergodicity becomes a potential
problem when there exist more than one regions of phase space that are
separated by some potential wall, because the ``kinetic energy'' of the
molecular dynamics system may not be enough to go through the potential wall.
This situation happens for QCD in four dimensions because of the non-trivial
topological sectors.
In the continuum theory the potential barrier is infinite and the gauge
configurations in different topological sectors cannot be reached by a single
stream of the continuous evolution.
On the lattice, the potential barrier is finite (order of inverse lattice
spacing $1/a$) and the probability of tunneling among different topological
sectors is non-zero, but will exponentially drop
($\sim e^{-U_0/a}$ with $U_0/a$ a nominal potential height) as the
continuum limit is approached. 
This means that the correct sampling of topological charge and thus the
valid simulation of the $\theta$-vacuum of QCD will become increasingly more
difficult\cite{Alles:1996vn, Del Debbio:2004xh}. 
In fact, this problem already manifests itself in the recent dynamical overlap
fermion simulations by the JLQCD collaboration
\cite{Kaneko:2006pa,Hashimoto:2006rb,Matsufuru:2006xr,Yamada:2006fr}, 
since they explicitly introduce a term that prevents topology change
\cite{Fukaya:2006vs} 
in order to avoid large numerical cost due to the
discontinuity of the overlap operator along the topology boundary.

One may then ask whether lattice QCD simulations with a fixed topological
charge are useful in general to extract physics of real world, {\it i.e.} QCD
at a given value of $\theta$. 
(If the CP symmetry is preserved, $\theta=0$.
Small but non-zero $\theta$ implies an interesting physics case to give rise
to the neutron electric dipole moment.)
The fixed topology simulation is desirable in the study of the
$\epsilon$-regime of QCD, since the physical quantities have striking
dependence on the topological charge, which is an important part of the physics
we are interested in.
(For a recent unquenched lattice simulation in the $\epsilon$-regime, see
\cite{Fukaya:2007fb,Fukaya:2007yv}.)
But in the $p$-regime, the fixed topology simulation gives rise to a
systematic effect.
In this paper we try to answer this question theoretically without relying on
any particular model or assumption. 
Extending a previous work by Brower {\it et al.} \cite{Brower:2003yx}, we show
that observables calculated at a fixed topological charge are different from
those at $\theta = 0$ by an amount of order of inverse space-time volume
$1/V$, so that both agree in the infinite volume limit. 

This result can be understood intuitively within the instanton picture of
the topological excitation in the QCD vacuum.
The instanton is a local object carrying a unit topological charge.
Starting from a trivial topological sector, an instanton and anti-instanton
can pair-create or -annihilate without changing the net topology and then may
apart from each other.
If we look at a region (or sub-volume) including the (anti-)instanton, that
region has locally non-trivial topological charge.
In this way, the topological fluctuation may occur even if the net topological
charge is kept fixed.
A supporting evidence of this expectation has been found in the chirality
density of low-lying eigenmodes of the overlap-Dirac operator, {\it i.e.}
local chirality is strongly peaked at $\pm 1$ even on the gauge configurations
with net topological charge $Q=0$ \cite{Hashimoto:2006rb}.
The effect of fixing the net topology will become relatively small as the total
volume becomes larger, as there is more chance to create (or annihilate)
the instanton-anti-instanton pairs.
The topological susceptibility $\chi_t$ characterizes the magnitude of these
local topological fluctuations.

In fact, in our formulation the topological susceptibility enters in the
difference of $O(1/V)$ between the fixed topological sector and the
fixed $\theta$ vacuum.
This type of finite volume effect can even be evaluated, once we know the
coefficient of $1/V$, which is found in general to be a second derivative of
the physical quantity of interest with respect to $\theta$, as well as the
topological susceptibility $\chi_t$.
In this paper, we describe the method to extract $\chi_t$ from a gauge
ensemble in the fixed topological sector, leaving the estimate of the finite
size effects for other physical quantities for a future publication.

This paper is organized as follows. 
In Section \ref{sec:general_formula}, by using the saddle
point expansion, we derive a formula which express the partition function 
at a fixed topological charge in terms of the partition function at 
$\theta=0$ and its derivatives with respect to $\theta$.
We extend this formula to the case for arbitrary correlation functions.
From these results it is easy to show that arbitrary correlation
functions at fixed topological charge agree with those at $\theta=0$ in
the infinite volume limit.
In Section \ref{sec:topological_susceptibility}, as an application of the
above formulas, we show that the topological susceptibility $\chi_t$ can be
extracted from two-point function of the topological charge density at a fixed 
topological charge. 
Through the Ward-Takahashi identities, we then relate the two-point function
of the topological charge density with a two-point function of the
pseudo-scalar density, which is more suitable for actual calculations with the 
overlap fermions. 
An extension to the case of four- and three-point correlation functions is also
discussed.
In Section~\ref{sec:CP-odd}, we consider a case of CP-odd observables taking
the calculation of the neutron electric dipole moment as an example.
Our conclusion is given in Section~\ref{sec:conclusion}.
In Appendix~\ref{sec:AppA} we check a validity of the saddle point expansion,
by comparing the expansion with an exact calculation for a simple example.
Results at order $V^{-3}$ are summarized in Appendix~\ref{sec:AppB}.

\section{General formula}
\label{sec:general_formula}

\subsection{Partition function at a fixed topological charge}
We first consider the partition function in the $\theta$ vacuum defined by
\begin{eqnarray}
  Z(\theta) &\equiv& \langle \theta \vert \theta \rangle 
  = \exp[-VE(\theta)], 
\end{eqnarray}
where $E(\theta)$ is the energy per (space-time) volume $V$.  
Without loss of generality we take $Z(0)=1$ as a normalization,
which is equivalent to $E(0)=0$. 
$E(\theta)$ satisfies the conditions
$E(\theta+2\pi)= E(\theta) = E(-\theta)$.
The topological susceptibility $\chi_t$ at $\theta=0$ is defined by
\begin{equation}
  \chi_t = \frac{\langle 0\vert Q^2\vert 0\rangle}{V},
\end{equation}
which is obtained by a second derivative of $E(\theta)$
\begin{equation}
  \chi_t =\left.\frac{d^2 E(\theta)}{d\theta^2} \right\vert_{\theta=0}.
\end{equation}
Since $\chi_t \ge 0$ by definition, $\theta =0$ is a local minimum of
$E(\theta)$. 
Moreover, Vafa and Witten proved that $Z(0) > Z(\theta)$ \cite{Vafa:1984xg}, 
which leads to $E(0) < E(\theta)$ for $\forall \theta \not=0$. 
Namely, $\theta =0$ is the global minimum of the function $E(\theta)$.
While we assume that $E(\theta)$ is analytic at $\theta=0$, 
$E(\theta)$ may have a non-analyticity at $\theta \not=0$, in particular at
$\theta=\pi$. 
For instance, the Chiral Perturbation Theory (ChPT) at the leading order gives 
\begin{equation}
  E(\theta) = \chi_t N_f^2 [1-\cos (\theta/N_f) ], \qquad -\pi < \theta \le \pi
\end{equation}
for $N_f$ the number of flavors.
A $\theta$ derivative of this particular form is discontinuous at
$\theta=\pm\pi$ for $N_f>1$ \cite{Witten:1980sp}. 
From these general conditions we can expand $E(\theta)$ around $\theta = 0$,
\begin{equation}
  E(\theta) = \sum_{n=1}^\infty  \frac{c_{2n}}{(2n)!} \theta^{2n} 
  = \frac{\chi_t}{2}\theta^2 + O(\theta^4).
\end{equation}

The partition function at a fixed topological charge $Q$ is a Fourier
coefficient of the periodic function $Z(\theta)$
\begin{equation}
  \label{eq:Z_Q}
  Z_Q = \frac{1}{2\pi}\int_{-\pi}^{\pi}\!d\theta\, Z(\theta) \exp(i\theta Q) 
  = \frac{1}{2\pi}\int_{-\pi}^{\pi}\!d\theta\,\exp( -V F(\theta)),
\end{equation}
where $F(\theta) \equiv E(\theta)-i\theta Q/V$,
and the Fourier expansion of $Z(\theta)$ is written as
\begin{equation}
  Z(\theta) = \sum_{Q=0,\pm 1,\pm 2, \cdots} Z_Q e^{-i\theta Q}.
\end{equation}
For a large enough volume, we can evaluate the $\theta$ integral in
(\ref{eq:Z_Q}) by the saddle point expansion.
The saddle point $\theta_c$ is given by
\begin{equation}
  \theta_c = i \frac{Q}{\chi_t V}(1+O(\delta^2)) 
\end{equation}
where $\delta\equiv Q/(\chi_t V)$.
We then expand $F(\theta)$ as
\begin{equation}
  F(\theta) = F(\theta_c) + \frac{E^{(2)}}{2} (\theta_c) (\theta - \theta_c)^2
  +\sum_{n=3}^\infty \frac{E^{(n)}(\theta_c)}{n!} (\theta - \theta_c)^n,
\end{equation}
where $E^{(n)}$ is the $n$-th derivative of $E(\theta)$ with respect to
$\theta$ at $\theta=\theta_c$, and is given by 
\begin{eqnarray}
  V F(\theta_c) &=& \frac{Q^2}{2\chi_t V}( 1+ O(\delta^2)), \\
  E^{(2)}(\theta_c) &=& \chi_t  ( 1+ O(\delta^2)), \\
  E^{(2n)}(\theta_c) &=& c_{2n} ( 1+ O(\delta^2)), \\
  E^{(2n-1)}(\theta_c) &=& \theta_c  c_{2n}(1+ O(\delta^2)).
\end{eqnarray}
Hereafter we omit an argument $\theta_c$ of $E^{(n)}(\theta_c)$ and
simply write it as $E^{(n)}$ unless otherwise stated.
By a change of variable $s = \sqrt{ E^{(2)} V }(\theta - \theta_c)$
we can rewrite the integral as
\begin{eqnarray}
  Z_Q &=&
  \frac{e^{-VF(\theta_c)}}{2\pi\sqrt{E^{(2)}V}}
  \int_{\sqrt{E^{(2)}V}(-\pi-\theta_c)}^{\sqrt{E^{(2)}V}(\pi-\theta_c)}
  ds\,  
  \exp\left[ -\frac{s^2}{2} - \sum_{n=3}\frac{E^{(n)}V}{n!}
    \left(\frac{s}{\sqrt{E^{(2)}V}}\right)^n\right] 
  \nonumber \\
  &=&
  \frac{e^{-VF(\theta_c)}}{2\pi\sqrt{E^{(2)}V}}
  \int_{-\infty}^\infty
  ds\,
  \exp\left[ -\frac{s^2}{2}-\sum_{n=3}\frac{E^{(n)}V}{n!}
    \left(\frac{s}{\sqrt{E^{(2)}V}}\right)^n\right]
  +O\left( e^{-V} \right)
  \nonumber \\
  &=&
  \frac{e^{-VF(\theta_c)}}{\sqrt{2\pi E^{(2)}V}}
  \left\langle \exp\left[ -\sum_{n=3}\frac{E^{(n)} V}{n!}
      \left(\frac{s}{\sqrt{E^{(2)}V}}\right)^n\right]
  \right\rangle + O\left( e^{-V} \right).
\end{eqnarray}
We defined
\begin{equation}
  \langle f(s) \rangle = \frac{1}{\sqrt{2\pi}}\int_{-\infty}^\infty ds\,
  e^{-s^2/2} f(s),
\end{equation}
with which $\langle s^{2n} \rangle  = (2n-1)!!$.
Neglecting exponentially suppressed terms and expanding in powers of $1/V$,
we obtain 
\begin{eqnarray}
  Z_Q &=& 
  \frac{e^{-VF(\theta_c)}}{\sqrt{2\pi E^{(2)}V}}
  \left[1 - 
    \left\langle\sum_{n=3}\frac{E^{(n)}V}{n!}
      \left(\frac{s}{\sqrt{E^{(2)}V}}\right)^n
    \right\rangle + 
    \cdots\right]
  \nonumber \\
  &=&\frac{e^{-VF(\theta_c)}}{\sqrt{2\pi E^{(2)}V}}
  \left[ 1 - \frac{E^{(4)}V}{8(E^{(2)}V)^2} + O\left(\frac{1}{V^{2}}\right) \right]
  \nonumber \\
  &=& \frac{1}{\sqrt{2\pi\chi_t V}}\exp\left[ - \frac{Q^2}{2\chi_t V}\right]
  \left[1 -\frac{c_4}{8 V \chi_t^2} + O\left(\frac{1}{V^{2}}, \delta^2 \right)
  \right].
\end{eqnarray}
This shows that, as long as $\delta\ll 1$ (equivalently, $Q \ll \chi_t V$), the
distribution of $Q$ becomes the Gaussian distribution.
Note however that the distribution can deviate from a Gaussian for
$Q=O(V)$, as numerically observed on a quenched lattice 
\cite{Del Debbio:2002xa, D'Elia:2003gr, Giusti:2007tu, Del Debbio:2007kz}.

\subsection{Correlation functions}
Here we consider an arbitrary correlation function
\begin{equation}
  G(\theta) = \langle \theta \vert  O_1 O_2 \cdots O_n \vert \theta\rangle, 
\end{equation}
whose Fourier coefficient at a fixed topological charge $Q$ is defined by 
\begin{equation}
  G_Q = \frac{1}{Z_Q}\frac{1}{2\pi}\int d\,\theta Z(\theta)\,
 G(\theta)\exp(i\theta Q).
\end{equation}
Note that the operators $O_i$ do not contain $\theta$, and the $\theta$
dependence comes solely from the vacuum angle.
Using the saddle point expansion as before, we obtain
\begin{eqnarray}
  G_Q &=& \frac{1}{Z_Q}
  \frac{e^{-VF(\theta_c)}}{ 2\pi\sqrt{E^{(2)}V}}\int_{-\infty}^{\infty}ds\,
  \exp\left[- \frac{s^2}{2} - \sum_{n=3}^\infty \frac{E^{(n)} V}{n!}
    \left(\frac{s}{\sqrt{E^{(2)}V}}\right)^n \right]
  G\left(\theta_c+\frac{s}{\sqrt{E^{(2)}V}}\right) 
  \nonumber \\
  &=& 
  G(\theta_c) + 
  \sum_{k=1}^\infty G^{(k)}(\theta_c) \frac{1}{k!}
  \left\langle\!\!\!\left\langle \left(\frac{s}{\sqrt{E^{(2)}V}}\right)^k
    \right\rangle\!\!\!\right\rangle,
\end{eqnarray}
where we define
\begin{equation}
  \langle\langle f(s) \rangle \rangle \equiv
  \frac{1}{Z_Q}
  \frac{e^{-VF(\theta_c)}}{2\pi\sqrt{E^{(2)}V}}\int_{-\infty}^{\infty}ds\,
  f(s) \exp\left[
    -\frac{s^2}{2} - 
    \sum_{n=3}^\infty \frac{E^{(n)}V}{n!}
    \left(\frac{s}{\sqrt{E^{(2)}V}}\right)^n \right]
\end{equation}
and $G^{(n)}$ is the $n$-th derivative of $G$ with respect to $\theta$.
Using the formulas 
\begin{eqnarray}
  \langle\langle 1 \rangle\rangle &=& 1, 
  \\
  \langle\langle  s^2 \rangle\rangle &=&
  1-\frac{E^{(4)}V}{2 (E^{(2)}V)^2} + \frac{5(E^{(3)}V)^2}{4(E^{(2)}V)^3}+
  O(V^{-2}), 
  \\
  \langle\langle  s^4 \rangle \rangle &=& 3!! + O(V^{-1}),
  \\
  \langle\langle s \rangle\rangle &=& 
  - \frac{E^{(3)}V}{2(E^{(2)}V)^{3/2}}
  \left(1-\frac{4E^{(4)}V}{3(E^{(2)}V)^2}+\frac{5(E^{(3)}V)^2}{4(E^{(2)}V)^3}\right)
  - \frac{E^{(5)}V}{8(E^{(2)}V)^{5/2}}+O(V^{-5/2})
  \\
  \langle\langle  s^3 \rangle \rangle &=&
  -\frac{5E^{(3)} V}{2(E^{(2)}V)^{3/2}} + O(V^{-3/2}).
\end{eqnarray}
we finally obtain
\begin{eqnarray}
  G_Q &=& 
  G(\theta_c) 
  + G^{(2)}(\theta_c) \frac{1}{2E^{(2)}V}
  \left(1 - \frac{E^{(4)} V}{2 (E^{(2)}V)^2}+\frac{5(E^{(3)}V)^2}{4(E^{(2)}V)^3}\right)
  + G^{(4)}(\theta_c)\frac{1}{8(E^{(2)}V)^2}
  \nonumber\\
  & & 
  - G^{(1)}(\theta_c)
  \left[\frac{E^{(3)} V}{2 (E^{(2)}V)^2} 
    \left(1-\frac{4 E^{(4)}V}{3(E^{(2)}V)^2}+\frac{5(E^{(3)}V)^2}{4(E^{(2)}V)^3}\right)
    +\frac{E^{(5)} V}{8(E^{(2)}V)^3}
  \right]
  \nonumber\\
  &&
  - G^{(3)}(\theta_c)\frac{5E^{(3)} V}{12(E^{(2)}V)^{3}} 
  + O(V^{-3}).
\end{eqnarray}
The above expansion is valid for any $\theta_c$ as long as 
$G^{(n)}(\theta_c)/G(\theta_c) =O(1)$ as $V\rightarrow \infty$.

Depending on the size of $\theta_c$, we can further expand the above formula.
If we take $\theta_c = O(V^{-1})$, (equivalently, $Q=O(1)$), we have
\begin{eqnarray}
  G_Q &=& G(0)+G^{(1)}(0)\theta_c 
  + G^{(2)}(0)\frac{\theta_c^2}{2}
  + G^{(2)}(0)\frac{1}{2E^{(2)}V}\left(1-\frac{E^{(4)}V}{2(E^{(2)}V)^2}\right) 
  + G^{(3)}\frac{\theta_c}{2E^{(2)}V}
  \nonumber \\
  &&
  + G^{(4)}(0)\frac{1}{8(E^{(2)}V)^2}
  - G^{(1)}(0)\left[\frac{E^{(3)} V}{2 (E^{(2)}V)^2} 
    \left(1-\frac{4 E^{(4)}V}{3 (E^{(2)}V)^2}\right)
    +\frac{E^{(5)} V}{8(E^{(2)}V)^3}\right]
  \nonumber\\
  &&
  - G^{(2)}(0)\theta_c\frac{E^{(3)} V}{2(E^{(2)}V)^2}
  + O(V^{-3})
  \nonumber \\
  &=& G(0) 
  + G^{(2)}(0)\frac{1}{2\chi_t V}
  \left[1-\frac{Q^2}{\chi_tV}-\frac{c_4}{2\chi_t^2V}\right]
  + G^{(4)}(0)\frac{1}{8\chi_t^2 V^2}
  \nonumber \\
  && 
  + G^{(1)}(0)\frac{iQ}{\chi_t V}\left(1-\frac{c_4}{2\chi_t^2 V}\right)
  + G^{(3)}(0)\frac{i Q}{2\chi_t^2 V^2}
  + O(V^{-3}).
  \label{eq:expansion}
\end{eqnarray}
If $G$ is CP-even, $G$ is an even function of $\theta$, so that
\begin{eqnarray}
  G_Q^{\mathrm{even}}
  &=& G(0) +G^{(2)}(0)\frac{1}{2\chi_t V}\left[1-\frac{Q^2}{\chi_tV}-\frac{c_4}{2\chi_t^2V}\right]
  + G^{(4)}(0)\frac{1}{8\chi_t^2 V^2} + O(V^{-3}), \nonumber \\
  \label{eq:even}
\end{eqnarray}
while, if $G$ is CP-odd, we have
\begin{eqnarray}
  G_Q^{\mathrm{odd}}
  &=&G^{(1)}(0)\frac{iQ}{\chi_t V}\left(1-\frac{c_4}{2\chi_t^2 V}\right)
  +G^{(3)}(0)\frac{i Q}{2\chi_t^2 V^2}
  +O(V^{-3}) .
\label{eq:odd}
\end{eqnarray}
In order to claim that the above expansion is convergent,
$G^{(n)}(0)/G(\theta)$ must be $O(1)$.
This condition is satisfied because the expansion
\begin{equation}
  G(\theta) = G(0) + \sum_{n=1}^\infty G^{(n)}(0) \frac{\theta^n}{n!}
\end{equation}
is valid for $\forall \theta =O(1)$.

The formula (\ref{eq:even}) provides an estimate of the finite size effect
due to the fixed topological charge.
The leading correction is of order $O(1/V)$ as advertised.
The dimension is compensated by the topological susceptibility $\chi_t$.
In ChPT it is evaluated as $\chi_t=m\Sigma/N_f$ by the sea quark mass $m$ and
the chiral condensate $\Sigma$ as well as the number of flavors $N_f$
\cite{Leutwyler:1992yt}.
The finite volume correction is suppressed when the quark mass is larger than
$1/(\Sigma V)$, while becomes significant when $m\sim 1/(\Sigma V)$.
This is consistent with the fact that the topological charge has a strong
effect in the $\epsilon$-regime, which is characterized by 
$m\Sigma V\lesssim 1$.
The correction has a coefficient $G^{(2)}(0)$ that represents the
$\theta$-dependence of the correlator.
This is not known in general but can be fitted with lattice data at various
$Q$.
In the mass region where ChPT is applicable, it can also be estimated, as done
in \cite{Brower:2003yx} at the tree level for pseudo-scalar meson mass.
One-loop calculations are in progress.

The other interesting formula (\ref{eq:odd}) suggests a possibility to
calculate CP-odd observables, such as the neutron electric dipole moment, at a
fixed non-zero topological charge, which will be discussed in
Section~\ref{sec:CP-odd}.

\section{Topological susceptibility}
\label{sec:topological_susceptibility}

In this section, we propose methods to extract the topological
susceptibility $\chi_t$ from correlation functions  
at a fixed topological charge $Q$.

\subsection{Two-point correlation function}
\subsubsection{Bosonic formula}
Suppose that there is a well-defined local operator $\omega(x)$ 
that measures the local topological charge.
The global topological charge $Q$ is then obtained as
$Q =\int d^4 x\,\omega(x)$, and the topological susceptibility is
$\chi_t = \int d^4 x \langle\omega(x)\omega(0)\rangle$, where the expectation
value is taken for the $\theta=0$ vacuum.

Since $\omega(x)\omega(0)$ is CP-even, Eq.~(\ref{eq:even}) gives
\begin{eqnarray}
  \label{eq:bosonic_two-point}
  \langle \omega(x) \omega(0) \rangle_Q &=&
  \langle \omega(x) \omega(0) \rangle +
  \langle \omega(x) \omega(0) \rangle^{(2)} 
  \frac{1}{2V\chi_t} 
  \left(1 - \frac{Q^2}{V\chi_t}-\frac{c_4}{2\chi_t^2 V}\right)
  \nonumber \\
  &+&\langle \omega(x) \omega(0) \rangle^{(4)} \frac{1}{8\chi_t^2V^2}
  +O(V^{-3}).
\end{eqnarray}
Here $\langle {\cal O}\rangle$ means a vacuum expectation value of
${\cal O}$ at $\theta=0$ and $\langle {\cal O}\rangle^{(n)}$ denotes its
$n$-th derivative with respect to $\theta$.
In the large separation limit $\vert x\vert\rightarrow\infty$, the CP
invariance at $\theta=0$ and the clustering property at a fixed $\theta$
\cite{Weinberg} gives 
\begin{equation}
  \label{eq:clustering}
  \langle \omega(x) \omega(0) \rangle \rightarrow
  \langle \omega(x) \rangle \langle \omega(0) \rangle = 0 .
\end{equation}
In addition,
\begin{equation}
  \langle \omega(x) \omega(0) \rangle^{(2)}
  = - \langle \omega(x) \omega(0) Q^2\rangle 
  +\langle \omega(x) \omega(0) \rangle \langle Q^2\rangle,
\end{equation}
whose second term vanishes as $\vert x\vert\rightarrow\infty$.
Denoting a connected VEV as $\langle \cdots \rangle_c$, 
the first term can be written as
\begin{eqnarray}
  && \int\!\!d^4y\,d^4z
  \langle \omega(x)\omega(0)\omega(y)\omega(z)\rangle
  = \int\!\!d^4y\,d^4z \Bigl[
  \langle \omega(x)\omega(0)\omega(y)\omega(z)\rangle_c 
  \nonumber \\
  &&
  + \langle \omega(x)\omega(0)\rangle \langle \omega(y)\omega(z)\rangle
  + \langle \omega(x)\omega(y)\rangle \langle \omega(0)\omega(z)\rangle
  + \langle \omega(x)\omega(z)\rangle \langle \omega(y)\omega(0)\rangle\Bigr] .
  \nonumber
\end{eqnarray}
As $\vert x\vert\rightarrow\infty$, the first and the second terms vanish and
we obtain 
\begin{equation}
  \int\!\!d^4y \int\!\!d^4z
  \langle \omega(x)\omega(y)\rangle \langle\omega(z)\omega(0)\rangle 
  + (y\leftrightarrow z)
  = 2
  \int\!\! d^4y \langle\omega(y)\omega(0)\rangle
  \int\!\! d^4z \langle\omega(x)\omega(z)\rangle.
\end{equation}
Using the translational invariance, this becomes $2\chi_t^2$.
Therefore, the term $\langle\omega(x)\omega(0)\rangle^{(2)}$ in
(\ref{eq:bosonic_two-point}) gives $-2\chi_t^2$ in the large separation
limit. 

Similarly, we consider 
\begin{equation}
  \langle\omega(x)\omega(0)\rangle^{(4)} =
  \langle\omega(x)\omega(0)Q^4\rangle
  - 6 \langle\omega(x)\omega(0)Q^2\rangle \langle Q^2\rangle 
  + \langle\omega(x)\omega(0)\rangle 
  \{ 6 \langle Q^2\rangle^2 -\langle Q^4\rangle \}.
  \label{eq:order4}
\end{equation}
The first term is written as
\begin{equation}
  \langle \omega(x) \omega (0) Q^4\rangle =
  \int \prod_i d^4x_i
  \langle\omega(x)\omega(0)\omega(x_1)\omega(x_2)\omega(x_3)\omega(x_4)\rangle,
\end{equation}
which may be decomposed in terms of the connected VEVs.
Since the terms containing both $\omega(x)$ and $\omega(0)$ in the same
connected VEV vanish, only the following terms remain:
\begin{eqnarray}
  && \int \prod_i d^4\, x_i \Bigl[
  8 \langle \omega(x) \omega(x_1)\omega(x_2)\omega(x_3)\rangle_c
  \langle \omega(0)\omega(x_4)\rangle
  \nonumber \\
  &&
  +12 \langle \omega(x) \omega(x_1)\rangle\langle\omega(x_2)\omega(x_3)\rangle
  \langle \omega(0)\omega(x_4)\rangle \Bigr]
  = 8\chi_t \frac{\langle Q^4\rangle_c}{V} + 12 V \chi_t^3.
\end{eqnarray}
Putting this into (\ref{eq:order4}) we have
\begin{equation}
  \label{eq:2pt-all}
  \langle \omega(x) \omega (0)\rangle^{(4)}
  \rightarrow 8 \chi_t \frac{\langle Q^4\rangle_c}{V}
  = \frac{8\chi_t}{V}\frac{d^4}{d \theta^4} \log Z(\theta) = - 8\chi_t c_4
\end{equation}
for the fourth-derivative term in (\ref{eq:bosonic_two-point}).

Gathering all terms we arrive at
\begin{equation}
  \label{eq:bosonic_two-pint_final}
  \lim_{\vert x\vert\to \mathrm{large}}
  \langle \omega(x) \omega(0) \rangle_Q =
  \frac{1}{V}\left(\frac{Q^2}{V}-\chi_t-\frac{c_4}{2\chi_t V} \right)
  +O(V^{-3}) + O(e^{- m_{\eta^\prime}\vert x\vert}),
\end{equation}
where the flavor singlet pseudo-scalar meson mass, $m_{\eta^\prime}$, is 
the lightest mass of possible intermediate states.  
This result explicitly shows that QCD at a fixed topological charge (in a
finite volume) is sick as a quantum field theory, because the clustering
property is violated. 
However, the magnitude of the violation can be estimated as a $1/V$ expansion
without introducing strong assumptions on the details of the dynamics of QCD.
The same result (up to the $c_4$ term) was obtained in a context of a
two-dimensional model in \cite{Fukaya:2004kp}.

Physical quantities such as the topological susceptibility $\chi_t$ can be
obtained through (\ref{eq:bosonic_two-pint_final}).
In practice, this formula will be used for a finite separation $x$ instead of
$|x|\to\infty$.
The clustering property (\ref{eq:clustering}) in the $\theta$ vacuum receives
a correction of order of $e^{-m_{\eta'}|x|}$, which vanishes quickly because
the flavor singlet meson $\eta'$ acquires a large mass due to the axial
anomaly of QCD.

\subsubsection{Fermionic formula}
We now express the bosonic correlation function
$\langle\omega(x)\omega(0)\rangle$ in terms of a fermionic one using
the anomalous axial U(1) Ward-Takahashi (WT) identities for 
an arbitrary operator $O$:  
\begin{equation}
  \label{eq:WT}
  \langle \partial_\mu A_\mu (x) O -  2m P(x) O +2\omega(x) O +
  \delta_x O \rangle = 0 ,
\end{equation}
where 
$A_\mu(x) =\frac{1}{N_f}\sum_f\bar\psi^f(x)\gamma_\mu\gamma_5\psi^f(x)$ 
and $P(x) = \frac{1}{N_f}\sum_f\bar\psi^f(x)\gamma_5\psi^f(x)$ are
the flavor singlet axial-vector current and pseudo-scalar density,
respectively,
and $\delta_x O$ denotes a axial rotation of the operator $O$ at $x$.
Here the quark field has a flavor index $f$ running from 1 to $N_f$.
The expectation value $\langle\cdots\rangle$ in (\ref{eq:WT}) can be taken
either in the $\theta$ vacuum or in the fixed $Q$ sector, since the WT
identities are valid with any external states.
Combining the following two WT identities 
(for $O=2mP(0)$ and $O=2\omega(x)$ with $x\not=0$)
\begin{eqnarray}
  \langle 2 m P(x) 2 m P(0) \rangle &=& 
  \langle \partial_\mu A_\mu (x)2 m P(0)\rangle + 
  \langle 2\omega (x) 2m P(0)\rangle 
  \\
  \langle 2\omega(x) 2 m P(0)  \rangle &=& 
  \langle 2\omega(x) \partial_\mu A_\mu (0)\rangle + 
  \langle 2\omega (x) 2\omega(0)\rangle,
\end{eqnarray}
we obtain a relation at a fixed topological charge
\begin{eqnarray}
  \label{eq:2mP2mP}
  \langle 2 m P(x)2 m P(0) \rangle_Q &=&
  \langle 2\omega (x) 2\omega (0) \rangle_Q +
  \langle \partial_\mu A_\mu (x) 2m P(0)\rangle_Q 
  + \langle 2\omega (x) \partial_\mu A_\mu (0)\rangle_Q.
\end{eqnarray}
In the large separation limit $\vert x\vert\rightarrow\infty$,
using (\ref{eq:even}) the second term becomes
\begin{eqnarray}
  \lefteqn{\langle \partial_\mu A_\mu (x) 2m P(0)\rangle_Q
  \rightarrow
  \langle \partial_\mu A_\mu (x)\rangle\langle 2 m P(0)\rangle}
  \nonumber \\
  &+&
  \frac{1}{2V\chi_t}\left(1-\frac{Q^2}{V\chi_t}-\frac{c_4}{2\chi_t^2V}\right)
  \langle \partial_\mu A_\mu (x) 2 m P(0)\rangle^{(2)}
  +\frac{1}{8\chi_t^2 V^2} \langle\partial_\mu A_\mu(x) 2mP(0)\rangle^{(4)}.
\end{eqnarray}
The first term on the right hand side vanishes, and the second term is
evaluated as 
\begin{equation}
  \langle \partial_\mu A_\mu (x) 2m P(0)\rangle^{(2)} =
  \langle \partial_\mu A_\mu (x) 2m P(0) Q^2\rangle
  \rightarrow 
  2 \langle \partial_\mu A_\mu (x) Q\rangle \langle 2m P(0) Q\rangle .
\end{equation}
Since the translational invariance (plus an appropriate boundary condition)
leads to 
\begin{equation}
  \langle \partial_\mu A_\mu (x) Q\rangle = \frac{1}{V}
  \int d^4\, x \langle \partial_\mu A_\mu (x) Q\rangle = 0
\end{equation}
we obtain $\langle \partial_\mu A_\mu (x) 2m P(0)\rangle^{(2)} =0$
in the large separation limit.

We next consider the fourth derivative term. In general,
\begin{equation}
  \langle A(x) B(0) \rangle^{(4)} =
  \langle A(x) B(0) Q^4\rangle -
  6\langle A(x) B(0) Q^2\rangle \langle Q^2 \rangle
  + \langle A(x) B(0) \rangle \{ 6 \langle Q^2\rangle^2-\langle Q^4\rangle \} 
\end{equation}
when $\langle A(x)\rangle = \langle B(0)\rangle = 0$. 
Using relations
\begin{eqnarray}
  \langle A(x) B(0) Q^2\rangle 
  &\rightarrow& 
  2\langle A(x) Q \rangle_c \langle B(0) Q\rangle_c, \\
  \langle A(x) B(0) Q^4\rangle 
  &=& 
  \langle A(x) B(0) Q^4 \rangle_c + \langle A(x) B(0)\rangle_c 
  \langle Q^4 \rangle_c 
  + 6 \langle A(x) B(0)Q^2\rangle_c  \langle Q^2 \rangle_c
  \nonumber \\ 
  &&
  +3\langle A(x) B(0)\rangle_c  \langle  Q^2 \rangle_c^2
  +4\langle A(x) Q\rangle_c  \langle B(0) Q^3 \rangle_c
  +4 \langle B(0) Q\rangle_c  \langle A(x) Q^3 \rangle_c
  \nonumber \\
  &&+12 \langle A(x) Q\rangle_c  \langle B(0) Q \rangle_c \langle Q^2\rangle_c
  \nonumber \\
  &\rightarrow&4 \langle A(x) Q\rangle_c  \langle B(0) Q^3 \rangle_c
  +4 \langle B(0) Q\rangle_c  \langle A(x) Q^3 \rangle_c
  \nonumber\\
  &&
  +12 \langle A(x) Q\rangle_c  \langle B(0) Q \rangle_c \langle Q^2\rangle_c,
\end{eqnarray}
the general formula reads
\begin{equation}
  \langle A(x) B(0) \rangle^{(4)} \rightarrow
  4 \langle A(x) Q\rangle_c  \langle B(0) Q^3 \rangle_c
  + 4 \langle B(0) Q\rangle_c  \langle A(x) Q^3 \rangle_c .
\end{equation}
For $A(x) = \partial_\mu A_\mu (x)$, the translational invariance
again gives $\langle A(x) Q\rangle_c = \langle A(x) Q^3\rangle_c =0$, and 
therefore
$\langle \partial_\mu A_\mu (x) 2 m P(0) \rangle^{(4)}\to 0$.
We thus conclude that the term
$\langle \partial_\mu A_\mu (x) 2 m P(0) \rangle_Q$
on the right hand side of (\ref{eq:2mP2mP}) vanishes
in the large separation limit $|x|\to\infty$, and the same conclusion also
holds for the last term $\langle 2\omega(x)\partial_\mu A_\mu(0) \rangle_Q$ in 
(\ref{eq:2mP2mP}).

Combining these we finally obtain
\begin{equation}
  \lim_{\vert x\vert\to \mathrm{large}}
  \langle m P(x) m P(0) \rangle_Q =
  \lim_{\vert x\vert\to \mathrm{large}}
  \langle \omega (x) \omega (0) \rangle_Q
  = 
  \frac{1}{V} \left(\frac{Q^2}{V}-\chi_t - \frac{c_4}{2 \chi_t V}\right) 
+ O(e^{-m_{\eta^\prime}\vert x\vert}).
  \label{eq:singlet}
\end{equation}
Namely, the topological susceptibility can also be calculated with the singlet
pseudo-scalar density correlator at a fixed topology.
If we apply the above evaluation to the non-singlet WT identities, we can show
\begin{equation}
  \lim_{\vert x\vert\to \mathrm{large}}
  \langle 2m P^a(x) 2m P^a(0) \rangle_Q =
  \lim_{\vert x\vert\to \mathrm{large}}
  \langle \partial_\mu A_\mu^a (x) 2mP^a (0) \rangle_Q 
  = 0,
  \label{eq:non-singlet}
\end{equation}
where $A_\mu^a(x) = \bar \psi(x) \gamma_\mu\gamma_5 T^a \psi(x)$ 
and $P^a =\bar\psi(x)\gamma_5 T^a\psi(x)$ are
the flavor non-singlet axial-vector current and pseudo-scalar density, 
respectively, with
$\tr\,T^a = 0$ and $\tr(T^a T^b) =\delta^{ab}/N_f$.
By subtracting this equation from (\ref{eq:singlet}), we may write
\begin{eqnarray}
  \lim_{\vert x\vert\to \mathrm{large}}
  \langle m P(x) m P(0) \rangle_Q^{\mathrm{disc}} &\equiv&
  \lim_{\vert x\vert\to \mathrm{large}}
  \left[
    \langle m P(x) m P(0) \rangle_Q -
    \langle m P^a(x) m P^a(0) \rangle_Q 
  \right] \nonumber \\
  &=&
  \frac{1}{V} \left(\frac{Q^2}{V}-\chi_t -\frac{c_4}{2\chi_t V}\right)
+O(e^{-m_\pi\vert x\vert}),
  \label{eq:disc}
\end{eqnarray}
where the superscript ``disc'' represents a disconnected contribution, 
and no summation is taken here for $a$.
Since the large separation limit of the non-singlet correlator
(\ref{eq:non-singlet}) is saturated only after a slowly dumping factor
$e^{-m_\pi|x|}$ with the pion mass $m_\pi$, the singlet correlator
(\ref{eq:singlet}) is preferable for a practical use. 

\subsection{Four-point correlation function}
Here we extend our analysis to a four-point function.

\subsubsection{Bosonic formula}
We consider a CP-even observable
\begin{eqnarray}
  \langle \omega(x_1)\omega(x_2)\omega(x_3)\omega(x_4)\rangle_Q & = &
  \langle \omega(x_1)\omega(x_2)\omega(x_3)\omega(x_4)\rangle 
  \nonumber\\
  &&+
  \langle \omega(x_1)\omega(x_2)\omega(x_3)\omega(x_4)\rangle^{(2)} 
  \frac{1}{2V\chi_t} \left(1-\frac{Q^2}{V\chi_t}-\frac{c_4}{2\chi_t^2 V}\right)
  \nonumber\\
  &&+
  \langle \omega(x_1) \omega(x_2)\omega(x_3) \omega(x_4) \rangle^{(4)}
  \frac{1}{8\chi_t^2V^2} + O(V^{-3})
\end{eqnarray}
We assume that four topological charge density operators are far apart from
any others 
($\vert x_i - x_j\vert \rightarrow$ large for $\forall i \not= \forall j$).
In this limit both
$\langle \omega(x_1) \omega(x_2)\omega(x_3) \omega(x_4) \rangle$ and
$\langle \omega(x_1) \omega(x_2)\omega(x_3) \omega(x_4) \rangle^{(2)}$
vanish and
$\langle \omega(x_1) \omega(x_2)\omega(x_3) \omega(x_4) \rangle^{(4)}$
becomes $24\chi_t^4$, taking the procedure used to obtain the analogous formula
(\ref{eq:2pt-all}) in the two-point function case.
We obtain
\begin{equation}
  \label{eq:4pt-bosonic}
  \langle \omega(x_1) \omega(x_2) \omega (x_3) \omega(x_4)\rangle_Q
  = 3\frac{\chi_t^2}{V^2}\left[1+\frac{1}{\chi_t^2 V}(c_4-Q^2\chi_t)\right]^2 
  +O(V^{-4}).
\end{equation}
In addition to the leading $O(1/V^2)$ term, we also give a next-to-leading
$O(1/V^3)$ correction.
Details of this $O(1/V^{3})$ calculation are given in Appendix~\ref{sec:AppB}.

\subsubsection{Fermionic formula}
Again we use the anomalous axial WT identities, which give a chain of equations
\begin{eqnarray}
  \langle P_1 P_2 P_3 P_4 \rangle &=& \langle dA_1 P_2 P_3 P_4\rangle +\langle \omega_1 P_2 P_3 P_4 \rangle \\ 
  \langle \omega_1 P_2 P_3 P_4 \rangle &=& \langle dA_2 \omega_1 P_3 P_4 \rangle + \langle \omega_1\omega_2 P_3 P_4 \rangle \\
  \langle \omega_1 \omega_2 P_3 P_4 \rangle &=& \langle dA_3 \omega_1 \omega_2 P_4 \rangle + \langle \omega_1\omega_2 \omega_3 P_4 \rangle \\
  \langle \omega_1 \omega_2 \omega_3 P_4 \rangle &=& \langle dA_4 \omega_1 \omega_2 \omega_3 \rangle + \langle \omega_1\omega_2 \omega_3 \omega_4 \rangle
\end{eqnarray}
where we have used a short-hand notation,
$P_i \equiv m P(x_i)$, $dA_i \equiv \partial_\mu A_\mu(x_i)/2$, 
and $\omega_i \equiv \omega (x_i)$. 
The operators are far apart from others as before.
Since the WT identities are valid in the fixed topological sector,
we obtain 
\begin{eqnarray}
  \langle P_1 P_2 P_3 P_4 \rangle_Q &=& 
  \langle \omega_1\omega_2 \omega_3 \omega_4 \rangle_Q\nn \\
  &&+
  \langle dA_1 P_2P_3P_4 + \omega_1 dA_2 P_3 P_4 + \omega_1\omega_2 dA_3 P_4
  +\omega_1\omega_2\omega_3 dA_4\rangle_Q .
\end{eqnarray}
From the clustering property and CP invariance the correlator
$\langle dA_i O_{jkl} \rangle$ vanishes, where 
$O_{jkl}\equiv F_j F_k F_l$ with $F_i = P_i$ or $\omega_i$
($i$, $j$, $k$, $l$ are all different).
In addition, $\langle dA_i O_{jkl} \rangle^{(2)}$ vanishes, because it is
written as 
$-\langle dA_i O_{jkl} Q^2\rangle \to 
 -\langle dA_i Q\rangle \langle O_{jkl} Q\rangle$.
Similarly, $\langle dA_i O_{jkl} \rangle^{(4)}$ vanishes, because it is
$\langle dA_i O_{jkl}  Q^4\rangle \to
4\langle dA_i Q\rangle \langle O_{jkl} Q^3\rangle$ and
$\langle \partial_\mu A(x) Q \rangle = (1/V)
  \int d^4 x\, \langle \partial_\mu A(x) Q \rangle = 0$.
We therefore obtain $\langle dA_i O_{jkl} \rangle_Q \rightarrow 0$,
which leads to
\begin{eqnarray}
\langle P_1 P_2 P_3 P_4 \rangle_Q &\rightarrow& 
\langle \omega_1\omega_2 \omega_3 \omega_4 \rangle_Q 
=
3\frac{\chi_t^2}{V^2}\left[1+\frac{1}{\chi_t^2 V}(c_4-Q^2\chi_t)\right]^2 
+O(V^{-4}).
\label{eq:singlet_4-point}
\end{eqnarray}

Similarly the non-singlet WT identities imply that
$\langle P_i^a P_j^a P_k P_l \rangle_Q$ and
$\langle P_i^a P_j^a P_k^b P_l^b \rangle_Q$ vanish
in the large separation limit, 
where $P_i^a = P^a(x_i)$ is a non-singlet PS density and 
$i$, $j$, $k$, $l$ are all different from others.
We then conclude that the disconnected correlator has the same asymptotic
form as $\langle\omega_1\omega_2\omega_3\omega_4\rangle_Q$:
\begin{eqnarray}
  \label{eq:4pt-fermionic}
  \langle P_1 P_2 P_3 P_4\rangle_Q^{\mathrm{disc}} 
  &\equiv&
  \langle P_1 P_2 P_3 P_4 \rangle_Q  
  - \{\langle P_1^a P_2^a P_3 P_4 \rangle_Q
  +\langle P_1 P_2 P_3^a P_4^a \rangle_Q
  -\langle P_1^a P_2^a P_3^b P_4^b \rangle_Q\}
  \nonumber\\
  && - \{ 2\leftrightarrow 3\} -\{ 2\leftrightarrow 4\}
  \nonumber\\
  & \to &
  3\frac{\chi_t^2}{V^2}\left[1+\frac{1}{\chi_t^2 V}(c_4-Q^2\chi_t)\right]^2 
  +O(V^{-4}),
\label{eq:disc_4-point}
\end{eqnarray}
where no sum is taken for $a\not= b$.

The formulas (\ref{eq:4pt-bosonic}) and (\ref{eq:4pt-fermionic}) provide
another method to extract the topological susceptibility from a fixed
topological sector. 
Since the contamination from the higher order effect in $1/V$ with the unknown
constant $c_4$ appears with a different coefficient from the two-point case 
(\ref{eq:bosonic_two-pint_final}) and (\ref{eq:singlet}), both $\chi_t$ and
$c_4$ can be extracted by combining the two- and four-point correlators.
Note that this calculation is free from short-distance singularities, because
the operators are explicitly put apart from others.
This is in contrast to the usual definition
$\chi_t=(1/V)\int d^4x \langle\omega(x)\omega(0)\rangle$ (at a fixed $\theta$).

\subsection{Three-point correlation function}
Topological susceptibility can be extracted also from CP-odd observables. 
The simplest case is given by a three-point function as
\begin{eqnarray}
  \lim_{|x_i-x_j|\to\mathrm{large}}
  \langle\omega(x_1)\omega(x_2)\omega(x_3)\rangle_Q
  &=& 
  \lim_{|x_i-x_j|\to\mathrm{large}}
  \langle\omega(x_1)\omega(x_2)\omega(x_3)\rangle^{(3)}
  \frac{iQ}{2\chi_t^2 V^2} + O(V^{-3})
  \nonumber\\
  &=&
  -\lim_{|x_i-x_j|\to\mathrm{large}}
  \langle\omega(x_1)\omega(x_2)\omega(x_3)Q^3\rangle
  \frac{Q}{2\chi_t^2 V^2} + O(V^{-3})
  \nonumber\\ 
  &=&
  -3\chi_t\frac{Q}{V^2}
  \left[1+\frac{7c_4}{6\chi_t^2 V}-\frac{Q^2}{3\chi_t V}\right] + 
  O(V^{-4}).
\end{eqnarray}
As in the cases of two- and four-point functions we may rewrite the above
result in terms of the fermionic quantities
\begin{eqnarray}
  \lefteqn{
    \lim_{|x_i-x_j|\to\mathrm{large}}
    \langle m P(x_1)  mP(x_2) mP(x_3)\rangle_Q
    = 
    \lim_{|x_i-x_j|\to\mathrm{large}}
    \langle m P(x_1)  mP(x_2) mP(x_3)\rangle_Q^{\mathrm{disc}} 
  }
  \nonumber\\
  &=&
  \lim_{|x_i-x_j|\to\mathrm{large}}
  \langle \omega(x_1) \omega(x_2) \omega(x_3)\rangle_Q 
  =
  -3\chi_t\frac{Q}{V^2}
  \left[1+\frac{7c_4}{6\chi_t^2 V}-\frac{Q^2}{3\chi_t V}\right] + 
  O(V^{-4}).
\end{eqnarray}
As expected, the three-point function is useful only when the net topological
charge is non-zero.

\section{CP-odd electromagnetic form factor}
\label{sec:CP-odd}
The three-point function considered above is an example of CP-odd observable.
In our formulation, as the formula (\ref{eq:odd}) implies, the CP-odd
observable (or its first derivative with respect to $\theta$) can be
calculated in a (non-zero) fixed topological sector.
Another interesting example of this class is the neutron electric dipole
moment. 

The CP-odd electromagnetic form factor $F_3(q^2)$ is related to the neutron
electric dipole moment $d_N$ as
\begin{equation}
  d_N = \lim_{q^2\rightarrow 0} \frac{F_3(q^2)}{2m_N},
\end{equation}
where $q$ is a momentum transfer.
It has been shown in \cite{Shintani:2005xg} that $F_3(q^2)$ can be extracted 
from the correlation functions
$\langle \bar N(p) J_\mu^{\rm EM}(q) N(p^\prime) Q \rangle$
and
$\langle \bar N(p) N(p) Q \rangle$, 
where $N(p)$ is a neutron interpolating field and $J_\mu^{\rm EM}$ is the
electromagnetic current. 
The momentum transfer is denoted by $q=p-p^\prime$.

Once $\chi_t$ is extracted from the methods explained in the previous section,
these neutron correlators can be extracted from at a fixed topological charge
using (\ref{eq:odd})
\begin{eqnarray}
\langle \bar N(p) J_\mu^{\rm EM}(q) N(p^\prime) \rangle_Q^{\rm odd} &=&
\langle \bar N(p) J_\mu^{\rm EM}(q) N(p^\prime) Q \rangle \frac{Q}{\chi_t V} + O(V^{-2}) \\
\langle \bar N(p) N(p) \rangle_Q^{\rm odd} &=&
\langle \bar N(p) N(p) Q \rangle \frac{Q}{\chi_t V} + O(V^{-2}),
\end{eqnarray}
where the superscript ``odd''  is understood that a CP-odd part of the
correlation function is considered. 

\section{Conclusions and discussions}
\label{sec:conclusion}
In this paper, we have derived general formulas which express arbitrary
correlation functions at a fixed topological charge $Q$ in terms of
the same correlation function (and its derivatives) in the $\theta$ vacuum.
As expected from the intuitive argument of local topological excitations, the
difference between the fixed $Q$ vacuum and the fixed $\theta$ vacuum
disappears in the large volume limit as $1/V$.
The relation is established only using fundamental properties of the quantum
field theory, such as the cluster decomposition principle.

These formulas open a new possibility to calculate physical quantities in the
lattice QCD simulations at a fixed topological charge.
This will become unavoidable as the continuum limit is approached,
irrespective of the lattice fermion formulation one employs,
as far as the algorithm is based on the continuous evolution of the gauge
field. 
(A proposal to avoid this limitation has recently been proposed
\cite{Golterman:2007ni} .
Its numerical feasibility is yet to be investigated.)

For the correlators of the local topological charge operator, the $1/V$
expansion is worked out for two-, three-, and four-point functions
to the second non-trivial order.
The local topological charge operator may either be a bosonic one or a
fermionic singlet pseudo-scalar operator.
At the leading non-trivial order the topological susceptibility $\chi_t$
appears as an expansion parameter, and a higher order parameter $c_4$ appears
at the second order.
In principle, these parameters can be determined by the lattice data.
The different correlators and different $V$ and $Q$ may be used to check the
results. 
This method is free from short-distance singularities, since the local
topological charge operators are put apart from others and no contact term
appears. 
Numerical calculation is in progress by the JLQCD collaboration on the gauge
configurations generated with dynamical overlap fermion
\cite{Kaneko:2006pa,Hashimoto:2006rb,Matsufuru:2006xr,Yamada:2006fr}.
Once these parameters are numerically obtained, they can be used as input
parameters for other physical observables.

The limitation of the formulas comes from the use of the saddle point
expansion.
It requires that the volume is large enough, $\chi_t V\gg 1$, that the local
topological fluctuation is in fact active.
It corresponds to the condition $\langle 0|Q^2|0\rangle\gg 1$ for that volume.
In addition, in order that the saddle point $\theta_c \simeq Q/(\chi_tV)$ can be
expanded around $\theta=0$,
the (fixed) topological charge $|Q|$ must be much smaller
than $\chi_t V=\langle 0|Q^2|0\rangle$.
Since $\chi_t$ is estimated in ChPT as $\chi_t=m\Sigma/N_f$ for $N_f$
flavors of sea quarks, the condition is $|Q|\ll m\Sigma V$.
This means that the system must be well apart from the $\epsilon$-regime 
($m\Sigma V\lesssim 1$).
This makes sense, because in the $\epsilon$-regime the physical quantities have
substantial dependence on the topological charge and can not be simply
expressed by the saddle point expansion.

Our general formulas can also be applied to other observables, such as
masses, decay constants, and matrix elements. 
Although the knowledge of the $\theta$ dependence of the observables is needed,
only a few derivatives with respect to $\theta$ are sufficient for small
$\theta$, which is physically most relevant.
Using the systematic $1/V$ expansion one can in principle extract those
derivatives to arbitrary finite order by looking at the $Q$ depenence.
Combining them with $\chi_t$, $c_4$, ..., 
the physical observables in the small $\theta$ vacuum can be reconstructed.
This is not surprising, because if we were able to determine all the
coefficients in $E(\theta)$ we could compute observables in the $\theta$
vacuum by a reweighting method. 
The point of our work is to present a practically feasible strategy 
in which the computational effort is drastically reduced. 
For the quantities well described by ChPT, estimation of the $\theta$ 
dependence based on the chiral lagrangian is possible.
Some results at the leading order were obtained in \cite{Brower:2003yx}, and we
are extending them to the next-to-leading order. 
This can also be used as an independent consistency check.

Once $\chi_t$ is extracted, we can calculate the first derivative of the
CP-odd quantities from simulations at a fixed non-zero topological charge
using (\ref{eq:odd}).
The most interesting such quantity in the context of lattice QCD calculation
is the neutron electric dipole moment. 
For this quantity, the calculation of the CP-odd form factor in the fixed
non-zero topological charge suffices to predict the physical result in the
$\theta$ vacuum.

\section*{Acknowledgments}
The authors would like to acknowledge the workshop at Yukawa Institute  
YITP-W-05-25 ``Actions and Symmetries in Lattice Gauge Theories,''
where part of this work was initiated.
S.A. would like thank Prof. K.F.~Liu for useful discussions.
This work is supported in part by the Grants-in-Aid for
Scientific Research from the Ministry of Education,
Culture, Sports, Science and Technology.
(Nos. 13135204, 15204015, 15540251, 16028201, 18034011, 18340075, 18840045,
19540286). 

\begin{appendix}
\section{Comparison of the saddle point expansion with an exact $\theta$ integration} 
\label{sec:AppA}
In this appendix we demonstrate that the saddle point expansion
reproduces the results from the exact $\theta$ integration, by considering
a simple model for $E(\theta)$. 

We consider the following form of the $\theta$ dependence:
\begin{equation}
  E(\theta) = \chi_t \left[ 1-\cos\theta\right].
\end{equation}
The partition function $Z_Q$ can be exactly calculated as
\begin{equation}
  Z_Q = e^{- V\chi_t}I_Q(V\chi_t),
\end{equation}
where $I_Q$ is the modified Bessel function.
If the form of $G(\theta)$ is given, $G_Q$ can be also calculated
exactly. 
For example, let us consider the case that  
$G(\theta)=\langle\theta|Q|\theta\rangle$ or $\langle\theta|Q^2|\theta\rangle$.
From $\langle\theta|Q|\theta\rangle=-iV dE/d\theta=-iV\chi_t\sin\theta$,
we have
\begin{equation}
  \langle Q \rangle_Q =
  \frac{-i}{Z_Q}
  \int_{-\pi}^\pi \frac{d\theta}{2\pi}
  V\chi_t \sin\theta\, e^{-VE(\theta) + i\theta Q}
  =V\chi_t \frac{I_{Q-1}(V\chi_t)-I_{Q+1}(V\chi_t)}{2I_Q(V\chi_t)}.
\end{equation}
Using a formula for the Bessel function 
$I_{n-1}(z)-I_{n+1}(z) = (2n/z)I_n(z)$,
the correct result $\langle Q \rangle_Q=Q$ is reproduced.
Similarly, using
\begin{equation}
  \langle\theta|Q^2|\theta\rangle
  =
  \frac{d^2}{d\theta^2}\log Z(\theta)
  +\langle\theta|Q|\theta\rangle^2
  = V\chi_t\cos\theta - (V\chi_t\sin\theta)^2,
\end{equation}
we obtain an expression
\begin{equation}
  \langle Q^2 \rangle_Q =  
  \frac{
    z      [ I_{Q+1}(V\chi_t)+I_{Q-1}(V\chi_t)] +
    (z^2/2)[ I_{Q+2}(V\chi_t)+I_{Q-2}(V\chi_t)-2I_Q(V\chi_t)]
  }{2I_Q(V\chi_t)},
\end{equation}
which leads to the correct result $\langle Q^2\rangle_Q =Q^2$.

We now consider a general $G(\theta)$. We will calculate $G_Q$, using the
expansion of $G(\theta)$
\begin{equation}
  G_Q = G(0) + \sum_{n=1}^\infty \frac{G^{(n)}}{n!} \langle \theta^n \rangle,
\end{equation}
where 
\begin{equation}
  \langle\theta^n\rangle_Q
  =
  \frac{1}{Z_Q}\int_{-\pi}^\pi \frac{d\theta}{2\pi} Z(\theta)e^{i\theta Q} 
  \theta^n .
\end{equation}
This expansion can be evaluated using the connected part
\begin{equation}
 \langle(i\theta)^n\rangle_{Q,c} = \frac{d^n}{dQ^n}\log Z_Q
\end{equation}
and the formulas
\begin{eqnarray}
  \langle X \rangle &=& \langle X \rangle_c, \\
  \langle X^2 \rangle &=& \langle X^2 \rangle_c+ \langle X \rangle^2, \\
  \langle X^3 \rangle &=& \langle X^3 \rangle_c+ 3\langle X^2 \rangle_c
  \langle X \rangle + \langle X \rangle^3,  \\
  \langle X^4 \rangle &=& \langle X^4 \rangle_c+ 4\langle X^3 \rangle_c
  \langle X \rangle + 3(\langle X^2 \rangle_c)^2 + 
  6\langle X^2 \rangle_c\langle X \rangle^2 + \langle X \rangle^4,
\end{eqnarray}
with $X=\theta$ in our case.
We evaluate the $Q$ derivatives in the large $V$ limit, using the asymptotic
expansion of the Bessel function
\begin{equation}
  I_Q(z) \simeq e^z\frac{1}{\sqrt{2\pi Z}}\sum_{n=0}^\infty
  (-1)^n (Q, n) \frac{1}{(2z)^n}+ O(e^{-z}),
\end{equation}
where $z=\chi_t V$ and
\begin{equation}
  (Q,n) = \frac{(4Q^2 -1^2)(4Q^2-3^2)\cdots (4Q^2-(2n-1)^2)}
  {n!\ 2^{2n}},
  \quad (Q,0)=1.
\end{equation}
Using these the partition function is written as
\begin{equation}
  Z_Q \simeq \frac{1}{\sqrt{2\pi z}}\left[1-\frac{(Q,1)}{2z}
    +\frac{(Q,2)}{(2z)^2}+\cdots \right]
\end{equation}
and we obtain ignoring $O(V^{-3})$ contributions
\begin{eqnarray}
  \log Z_Q & = &-\log \sqrt{2\pi z} -\frac{1}{2z}
  \left[(Q,1)+\frac{(Q,1)^2}{2(2z)}-\frac{(Q,2)}{2z} \right] \nonumber \\
  &=&
  -\log \sqrt{2\pi z}
  -\frac{1}{2z}\left(Q^2-\frac{1}{4}\right)\left(1+\frac{1}{2z}\right) , 
\end{eqnarray}
from which
\begin{eqnarray}
  \frac{d}{dQ}\log Z_Q &=& -\frac{Q}{z}\left(1+\frac{1}{2z}\right),\\
  \frac{d^2}{dQ^2}\log Z_Q &=& -\frac{1}{z}\left(1+\frac{1}{2z}\right),\\
  \frac{d^n}{dQ^n}\log Z_Q &=& 0, \quad n=3,4,5,\cdots .
\end{eqnarray}
Therefore, 
\begin{eqnarray}
  \langle \theta \rangle_Q &=& i\frac{Q}{z}\left(1+\frac{1}{2z}\right), \\
  \langle \theta^2\rangle_Q &=&
  \frac{1}{z}\left(1+\frac{1}{2z}\right)-\frac{Q^2}{z^2}+O(z^{-3}),\\
  \langle \theta^3\rangle_Q &=& i\frac{3Q}{z^2}+O(z^{-3}),\\
  \langle \theta^4\rangle_Q &=& \frac{3}{z^2}+O(z^{-3}) .
\end{eqnarray}
Collecting these the final result becomes
\begin{eqnarray}
  G_Q &=& G(0)
  +G^{(2)}(0)\frac{1}{2z}(1+\frac{1}{2z}-\frac{Q^2}{z}) 
  +G^{(4)}(0)\frac{1}{8 z^2}\\
  &&
  +G^{(1)}(0) i\frac{Q}{z}(1+\frac{1}{2z}) 
  +G^{(3)}(0)i\frac{Q}{2z^2}
\end{eqnarray}
Noticing that $c_4 = -\chi_t$ for the present case,
this result completely agrees with the previous result (\ref{eq:expansion})
within the $O(V^{-3})$ errors.
This demonstrates that the saddle point expansion reproduces correct 
results, up to exponentially small corrections.

The leading order of the chiral perturbation theory gives
\begin{equation}
  E(\theta) = N_f^2 \chi_t \left[ 1-\cos(\theta/N_f)\right] .
\end{equation}
In this case, we can
extend the integration range to $-N_f\pi < \theta < N_f \pi$  
by ignoring exponentially small contribution.
After performing $\theta$ integration exactly, we obtain
\begin{equation}
  Z_Q = e^{- z} I_{\hat Q}(z) + O(e^{-z [1-\cos (\pi/N_f)]})
\end{equation}
where $z= V N_f^2 \chi_t$ and $\hat Q = N_f^2 Q$.
Noticing that $c_4= -\chi_t/N_f^2$ in this case,
it is easy to see that the saddle point
expansion agrees with exact results order by order in the $1/V$
expansion, as long as exponentially small corrections are ignored.

\section{Contributions at $V^{-3}$}
\label{sec:AppB}

\subsection{CP-even contributions}
For $\theta_c=O(V^{-1})$, CP-even contributions become
\begin{eqnarray}
G_Q^{\rm even} &=&G+G^{(2)}\frac{\theta^2_c}{2}+
G^{(2)}\frac{1}{2x}\left(1-\frac{E^{(4)}V}{2x^2}
-\frac{E^{(6)}V}{8x^3}+\frac{2(E^{(4)}V)^2}{3 x^4}
+\frac{5(E^{(3)}V)^2}{4x^3}\right)\nn\\
&+&G^{(4)}\frac{\theta_c^2}{4 x}
+ G^{(4)}\frac{1}{4!}
\left(\frac{3}{x^2}
-\frac{4E^{(4)}V}{x^4}\right) + G^{(6)}\frac{15}{6!x^3}
-G^{(2)}\theta_c\frac{E^{(3)}V}{2x^2} + O(V^{-4}).
\end{eqnarray}
Using the large separation limit $G=G^{(2)}=0$, $G^{(4)}=24\chi_t^4$ and $G^{(6)}=480\chi_t^3 c_4$
for $G=\langle \omega(x_1)\omega(x_2)\omega(x_3)\omega(x_4)\rangle$,
we obtain
\begin{equation}
\langle \omega (x_1) \omega(x_2)\omega (x_3)\omega (x_4) \rangle_Q
\rightarrow 
\frac{3\chi_t^2}{V^2}\left[1 + \frac{2}{\chi_t^2
          V}(c_4-Q^2\chi_t)\right]
+O(V^{-4}) .
\end{equation}

\subsection{CP-odd contributions}
For $\theta_c = O(V^{-1})$, CP-odd contributions become
\begin{eqnarray}
G_Q^{\rm odd} &=&
G^{(1)}\theta_c +
G^{(3)}\frac{\theta_c^3}{3!}+G^{(3)}\frac{\theta_c}{2x}\left(
1-\frac{E^{(4)}V}{2x^2}\right)+G^{(5)}\frac{3\theta_c}{4! x^2}\nn \\
&-& G^{(1)}\left[\frac{E^{(3)}V}{2x^2}\left(1-\frac{4
           E^{(4)}V}{3x^2}\right)
+\frac{E^{5}V}{8x^3}\right] -G^{(3)}\frac{5 E^{(3)}V}{12x^3} .
\end{eqnarray}
Using the large separation limit $G^{(1)}=0$, $G^{(3)}= i 6 \chi_t^3$ and
$G^{(5)}= i60 \chi_t^2 c_4$
for $G = \langle \omega(x_1) \omega(x_2) \omega(x_3) \rangle $,
we obtain
\begin{equation}
\langle \omega (x_1) \omega(x_2)\omega (x_3) \rangle_Q \rightarrow
-\frac{Q}{V^2}\left[3\chi_t + \frac{7 c_4}{2\chi_t
           V}-\frac{Q^2}{V}\right].
\end{equation}

\end{appendix}
%


%
\end{document}